# Resolving duration and sequence of Na and continuum flashes during MBSL with time-correlated single photon counting

*T.V. Gordeychuk[*], M.V.Kazachek*

V.I.Il'ichev Pacific Oceanological Institute, Russian Academy of Sciences, Far East Branch, Vladivostok, Russia
[*]tanya@poi.dvo.ru

ABSTRACT

Time-correlated single photon counting (TCSPC) was used for the first time for resolving a duration and a sequence of the flashes of Na and continuum during MBSL from argon gas saturated aqueous NaCl and SDS solutions. It was found that a bubble giving both kinds of the flashes emits them simultaneously with a negligible delay of about 0.5 ns for Na flash. The duration of Na flash was defined as 4.6 ns, 7.5 ns, 11.2 ns for 5 M NaCl, 0.5 M NaCl, 5 mM SDS solutions, respectively. The duration of continuum flash was varying as 1.6-2.4 ns for all investigated solutions. To explain relatively large duration of Na flash, the scheme of evolution of presumable Na-emission zones is proposed based on the existence of an optimal temperature for Na emission.

## 1. Introduction

Sonoluminescence (SL) - a weak light emission, resulting from the high energy focused by a rapid and violent collapse of a cavitating bubble, which generates extremely high temperatures and pressures inside the bubble. SL can be seen from clouds of bubbles (multibubble sonoluminescence - MBSL) or from a stable pulsating single bubble levitated in a partially degassed liquid (single bubble sonoluminescence - SBSL). Regardless to SBSL or MBSL, SL occurs as short-time flashes correlating with a moment of bubble collapse. A duration and a shape of photons' packets, as well as a sequence of the emitting of different spectral components, are the important data for understanding SL origin. The three methods were used to estimate the duration of SL flashes. The direct estimations using oscilloscopes were represented for MBSL in [1-6] and for SBSL in [3]. The streak camera was used for SBSL in [7-10]. In [11, 12] the time-correlated single-photon counting (TCSPC) was successfully used for measuring the duration of SBSL flash.

The data relating to the duration and intercomparison of Na and continuum flashes during MBSL are represented in [3,4,6]. In [3] the flash duration was estimated with the oscilloscope as 10 ns and 70 ns for high and low ultrasonic power for MBSL from ethylene glycol NaCl solution saturated with Ar. In [4] it was found that for aqueous NaCl solution the timing of Na-emission pulses shows the random behavior and is wider in comparison with the pulses of continuum emission. The results which presented in [6] for the aqueous NaCl solution saturated with Xe allow to conclude that the time interval between Na flashes and continuum flashes within a single ultrasound period can exceed 25 ns. In [13] simultaneous measurements of continuum and Na flashes were made for the bubble created by laser. It was found that Na emission outstrips the continuum emission by several tens of nanoseconds. The data containing the information about a sequence of Na and continuum flashes for MBSL were not found in literature.

A complexity to obtain similar experimental data for multibubble system occurs mainly because the flashes from separate bubbles are not coincide in time and, besides, they are formed by the small number of photons. TCSPC have long been used successfully for the analysis of random radioactive decay events in nuclear physics. Therefore, the method seems to be suitable for extraction data about flashes emitted by one "statistically average" bubble within a multibubble system.

We examine TCSPC for studying MBSL from aqueous NaCl and SDS solutions. It is well known that MBSL spectra obtained from rare gas saturated aqueous solution containing metal ions consist of a broad continuum encompassing the UV/Visible range, OH-radical emission peak around 310 nm and emission lines from the excited states of metal atoms overlapping the continuum. The continuum, as suggested, formed with bremsstrahlung, blackbody radiation, recombination radiation and/or excited-state molecular emission. The appearance of atomic emission in SL spectra leaves many questions unanswered.

Here we attempt using TCSPC for extracting the emission from one "statistically average" bubble and estimating the duration of Na-atom flash and the continuum flash and their sequence for such a bubble within one act of SL emitting.

## 2. Experimental details

The experimental setup for MBSL spectra measuring is described elsewhere [14,15]. According to the aim of this work, the setup was improved using a correlation counter based on a digital oscilloscope RIGOL DS1104Z, a computer, and the author's software [16]. The scheme is shown in Fig. 1. Aqueous solutions of NaCl and sodium dodecyl sulfate (SDS) sparged with Ar 2 h before the irradiation and in the course of the work in 0.5 l glass reservoir placed in refrigerated circulator Julabo F12. Solutions were constantly rotated through a thermostated stainless steel irradiation cell (inner diameter 2 cm, volume 50 ml) with the help of a peristaltic pump (flow rate was ~ 1 ml/s). The temperature of solutions inside the cell during irradiation was constant at 10 ºC. The irradiation was conducted at ultrasound frequency of 20 kHz with the total absorbed power of 18 W measured by direct generator VC-750readings. The distance between a tip of a piezoceramic transducer (tip diameter was 13 mm) and a quartz window was 100 mm.

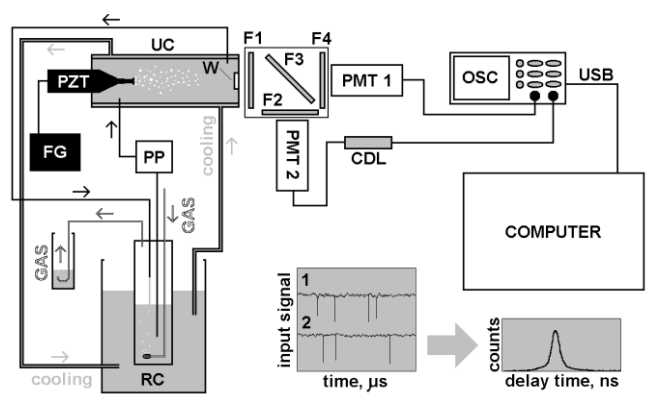

**Fig. 1.** Scheme of experimental setup: FG - frequency generator, PZT - transducer, PP - peristaltic pump, RC - refrigerated circulator, W - quartz window, F1, F2, F3, F4 - seats for optical filters, PMT1, PMT2 - photomultiplier tubes, CDL - constant delay line, OSC - digital oscilloscope.

Optical filters were used for dividing light emission and for separating different spectral regions. The filters could be installed in the positions F1-F4 in different combinations (Fig. 1). The flat surface of the filter in the position F3 acted as a semitransparent mirror dividing light flux between two photomultiplier tubes (PMT1 and PMT2). The range of PMTs sensitivity is 200-750 nm. The area of connection of the quartz window, the filters and the photomultipliers was carefully protected from outside light.

For measuring the autocorrelation functions between photon pulses which should be representative of the same type of emission ("Na-Na" or "continuum-continuum") the PMTs were simultaneously irradiated with light of the same spectral range. In this case, the suitable filter was installed in the position F1: the "red" filter with 550 nm cut-off wavelength for separating Na emission (Na D-line ~590 nm) or "blue" filter with the bandwidth of 350-430 nm for separating the continuum emission. The neutral filter was installed in the position F3 whereas the positions F2, F4 were empty. The transmittance spectra of the filters as compared to MBSL spectra from investigated solutions, measured at the low spectral resolution, are as shown in Fig. 2. Since Na emission provides a key contribution to spectral intensity within the "red" filter bandwidth we can talk only about Na emission concerning this part of the spectra. For measuring the correlation functions between photon pulses of different types of emission ("continuum-Na" and "Na-continuum") the red and blue filters were installed in the positions F2 and F3, respectively, or in reverse order.

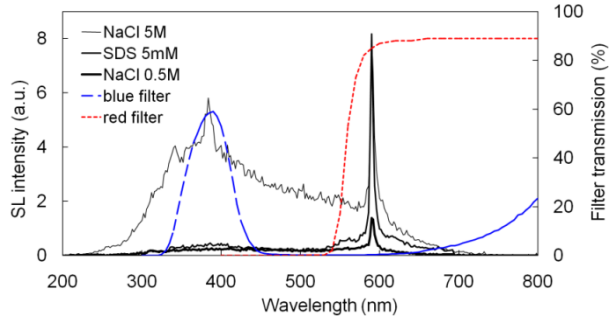

**Fig. 2.** Low resolution (2.9 nm FWHM) MBSL spectra from Ar saturated aqueous 5 M NaCl, 5 mM SDS, 0.5 M NaCl solutions and transmittance spectra of red and blue filters.

A constant delay line (6-m-long cable) was added between the PMT2 and the oscilloscope input, which provided a delay of approximately 30 ns. The delayed-coincidence method was thus implemented. The PMTs operated in a single-photon counting mode. The signals had been sent to two input channels of the oscilloscope. The adjustable 50-Ohm load resistors were connected in parallel with the oscilloscope inputs. The signals arriving from the PMTs at the oscilloscope inputs were the negative pulses with 2-5 mV amplitude, 4-8 ns duration and had a steep leading edge with duration of less than 1 ns. The frequency of the pulses depended on SL intensity and could vary from 100 Hz to 100 kHz. The frequency of dark counts was in the range of 100-300 Hz. The choice of a channel for the oscilloscope triggering was not important because the oscilloscope's recording length was enough for detecting the pulses from the both channels. The oscilloscope was operated with the computer equipped with the author's software defining the thresholds, counting of the pulses and calculation of correlations between the pulses. The software was written using "Visual Basic" implemented in Microsoft Excel.

An algorithm for finding the dependence between the flows of the pulses from two oscilloscope channels is as follows. A correlation function $k_{12}(\tau)$ was calculated for time segments of signals $f_1(t)$ and $f_2(t)$ read from the oscilloscope. Here $f(t)$ is an amplitude of the signal and the subscripts 1,2 means the oscilloscope channels. The segments were the arrays with length $M$=60000 pts. The length of the point, which is equal to a digitization time, was 2 ns in our case, so the length of each segment was 120 μs. We set a threshold $g$ for each channel and a requirement for the existence of a pulse at a time point $m$ as follows: the pulse exists if $f(t_{m-1}) \geq g$ and $f(t_m) < g$. Then we set a multiplication $f_1 \cdot f_2 = 1$ if the pulses coincide and $f_1 \cdot f_2 = 0$ if otherwise. The correlation function for discrete values of $\tau$ and $t$ is defined as

$$k_{12}(\tau) = \sum_{m=1}^{M} f_1(t_m) \cdot f_2(t_m + \tau) \qquad (1).$$

Passing over all $m$ values from 1 to $M$ in a loop and calculating the intervals $\tau$ between successive pulses in different channels, we fill in the arrays $k_{12}(\tau)$ and then read next signal segments $f_1(t)$ and $f_2(t)$. The function $k_{12}(\tau)$ is displayed after $f_1(t)$ and $f_2(t)$ have been reading the specified number of times. The $k_{12}(\tau)$ function gives the distribution of the number of events that consist of "a pulse in the first channel followed by a pulse in the second channel" over the time interval $\tau$ between the pulses. The developed algorithm has just one loop for each $f_1(t)$&$f_2(t)$ data treatment and takes time of $o(M)$. It provides an advantage over the algorithm based on the direct multiplication $f_1 \cdot f_2$ in accordance with (1), which requires the double loop for $m$ and $\tau$ and takes time of $o(M^2)$. In our case, the $k_{12}(\tau)$ calculation time was an order of magnitude less than the time of signal transfer via the USB cable because of the USB1.0 format supported by the oscilloscope. Therefore, the disadvantage is low time efficiency due to the delay in the oscilloscope–computer communication line, but it is not essential for studying processes for which the acquisition time is not critical. The algorithm had been tested with a random number generator. RIGOL DG4062 pulse generator and LEDs were used to test the setup with light pulses. The width of the correlation function was in well accord with the duration of LEDs flash (10-16 ns) [16]. To check a width of "own" correlation function of the counter, when delivering synchronous pulses, the pulses from PMT were applied to the oscilloscope input while the second input was connected to the first input by the constant delay line. The width calculated from Gaussian fitted curve was about 1.7 ns and was due to digitization.

The ideas on the usefulness of TCSPC for MBSL studying are as follows. Let us posit a cloud of a 1000 bubbles, each of which emits 1000 photons during one flash, thus in total $10^6$ photons. The number of photons in a flash was estimated as $\sim 10^6$ for SBSL [17]. Let the flash duration be about 50 ns. Let also the flashes occur randomly and be evenly distributed within the period of ultrasound (50 μs for 20 kHz frequency). The last assumption is based on the fact that MBSL flashes occupy most of a period [1,4]. Suppose that we can fix only one thousandth from all this flow in the form of the PMT's pulses, owing to the setup aperture, the quantum efficiency of the photocathode and the losses on the optical surfaces, thus in total 1000 pulses.

It is clear that there is a 1-in-1000 chance of two fixed pulses arriving from the same bubble than from two different bubbles. Consequently, among 1000 received correlations we have 1 from the same bubble and 999 from the different bubbles. The correlation of pulses related to "the same bubble" should be located within the delay range 0-50 ns, while the correlations of pulses from "the different bubbles" should be distributed over the delay range 0-50μs. That means that the correlation function should be a peak with 50 ns width and twice as high as a background. If the flash is shorter than 50 ns the efficiency of TCSPC is being enhanced because the peak of correlation function will be narrower and higher. We especially note that, in fact, we observe the correlation peak with the amplitude about 100 times more than the background!

## 3. Results

The screenshots from the oscilloscope with the image of MBSL pulses from Ar saturated aqueous 5 mM SDS solution incoming simultaneously from both PMTs to two oscilloscope channels are shown on Fig. 3. The image demonstrates the different delay distributions of pulses when the both PMTs are illuminated with the same spectral region: 350-430 nm region corresponds to the continuum (Fig. 3a), 550-750 nm region includes the Na line (Fig. 3b). One can see that the Na pulses are located within the wider delay distribution compared to the continuum pulses.

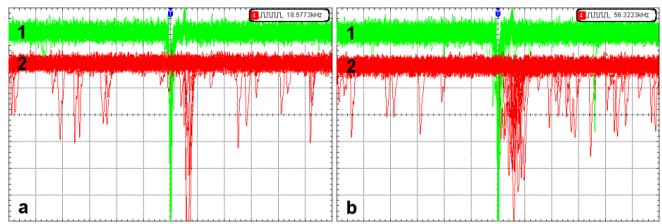

**Fig. 3.** Multiple screenshots from oscilloscope illustrating different delay distributions of photon pulses during MBSL from Ar saturated aqueous 5 mM SDS solution: (a) pulses of continuum and (b) pulses of Na emission. Channels of oscilloscope are marked as 1 (green) and 2 (red). The horizontal scale is 50 ns/div. Synchronization trigger is set to channel 1.

The autocorrelation functions for Na and continuum pulses are shown in Fig. 4. The FWHM of the function peak is associated with the duration of the flash from one "statistically average" bubble. The autocorrelation function background having low amplitude is formed by the pulses separated by time, which come from different bubbles. The FWHMs and the peak positions were calculated by fitting a Gaussian to the function profile. Taking into account that the width of convolution of two Gaussian functions is $\sqrt{2}$ more the width of original functions, the FWHM was divided into $\sqrt{2}$. The FWHM of own function of the counter (1.7 ns) was subtracted from the received value. The results show that the duration of the continuum flash was not significantly different for all investigated solutions, but was markedly less than the duration of Na flash for all cases. The duration of Na flash was different for all solutions (Table 1).

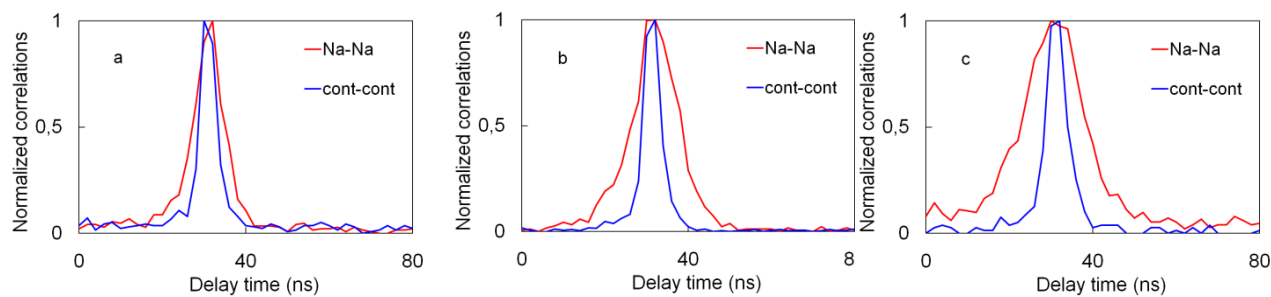

**Fig. 4.** Normalized autocorrelation functions of photon pulses of Na and continuum emission during MBSL from Ar saturated aqueous solutions (a) 5 M NaCl, (b) 0.5 M NaCl, (c) 5 mM SDS.

**Table 1.** Duration of Na and continuum flashes and "continuum-Na" delay during MBSL from Ar saturated aqueous NaCl and SDS solutions.

|  | NaCl, 5 M | NaCl, 0.5 M | SDS, 5 mM |
|---|---|---|---|
| Na | 4.6 ns | 7.5 ns | 11.2 ns |
| continuum | 1.6 ns | 1.6 ns | 2.4 ns |
| delay | 0.6 ns | 0.4 ns | 0.6 ns |

The correlation functions of "continuum-Na" and "Na-continuum" pulses are presented in Fig. 5. The functions were calculated from the pulses measured under transposition of optic filters in the positions F2 and F3 (Fig. 1). The shift between the peaks of the functions in Fig. 5 indicates the double delay between Na and continuum pulses (Table 1). The shift was regularly observed being about 1 ns. Since the duration of the flashes is significantly longer than the shift, one can concludes that Na and continuum flashes are overlapping. Briefly, the results indicate that Na and continuum emit almost simultaneously with negligible delay of about 0.5 ns for Na emission.

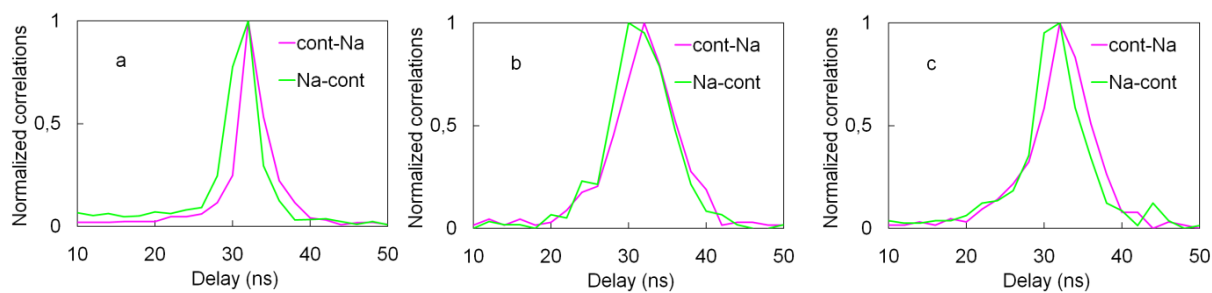

**Fig. 5.** Normalized correlation functions of photon pulses of Na and continuum emission during MBSL from Ar saturated aqueous (a) 5 M NaCl, (b) 0.5 M NaCl, (c) 5 mM SDS solutions.

## 4. Discussion

As noted above, the appearance of metal atomic emission in MBSL spectra from aqueous alkali-metal salt solutions remains unresolved issue. The mechanisms of ion reduction with subsequent metal-atom electronic excitation as well as the location of atomic emission are still in debate. Two different models have been proposed to explain the entrance nonvolatile alkali-metal ions into presumptive emission zones. The shell model suggests that the emission from alkali-metal atoms takes place in the thin liquid layer directly on the bubble surface where the ions undergo reducing and exciting through the reactions with radicals generated during the bubble collapse. The injected droplet model is also adopted under which the emission originates from the hot gas phase inside a bubble while alkali-metal ions enter the bubble with nanodroplets of the liquid before or after the collapse. The liquid droplets can form owing to the developing of surface instability and/or upon the coalescence and fragmentation of bubbles within the bubbles' cloud. Further, the high temperature causes thermolysis of salt with subsequent formation of electronically excited metal atoms.

Alkali-metal lines in MBSL spectra have a complex profile, different from the spectra in a flame. The Na and K lines consist of two components overlapping each other. An unshifted narrow component can be emitted from low-density exciting environment only, and that is unclear. A broadened asymmetrically shifted component was explained by integration of the spectra emitted in a wide density range [18]. In [19] was experimentally shown that a blue satellite of alkali-metal line and, consequently, a broad component emit from gas phase inside the bubble. The mechanism was also proposed including the reactions with H and OH radicals such as $OH+H+Na \rightarrow H_2O+Na^*$. The H and OH radicals appear during the final stage of bubble collapse when almost adiabatic rapid compression leads to the formation of "hot-spot" with the extreme temperatures and densities. A few papers presented experimental observations of the spatial separation of MBSL bubbles emitting the continuum and the metal line. The observation resulted in the division of cavitation bubbles into sonoluminescent bubbles which emit the continuum and chemiluminescent ones when alkali-metal emission is thought mostly occurs from chemical reactions than due to the thermal excitation. As illustrated in [20] the areas of chemiluminescent bubbles (an active yield of OH radicals) coincide with the areas of Na luminescence within the bubbles cloud. In addition, it was shown in [21] that luminescent bubbles were localized within the high intensity area close to the radiative surface of ultrasonic transducer. At the same time, chemiluminescent bubbles emitting Na line were observed along the streamers outside of the intense ultrasonic area. A similar effect was observed in [4]. The results were explained, in particular by more stable bubble surface when the bubble oscillates inside a dense bubbles cloud and it leads to the suppressing of the mechanism of nanodroplet injection.

On the other hand, Na emission was observed from a "stable" luminescing bubble in single bubble experiments [3,22,23]. There is also indirect indicators that at least in aqueous SDS solutions excited Na atoms should form directly at the bubble/solution interface in order to emit [20, 24]. The authors of [24] showed that the addition of $Zn^{2+}$ ions to SDS solution suppressed the Na line in MBSL spectra. The authors reasonably concluded that the replacement of $Na^+$ ions by $Zn^{2+}$ ions directly on the negatively charged surface of a bubble should not change the solution concentration of free $Na^+$ ions immediately near the bubble/solution boundary and should not affect the $Na^+$ ions injection with nanodroplets. Note that the conditions on the bubble/solution boundary are different for SAS and alkali-metal halides solutions. Sodium chloride is characterized by negative adsorption with respect to water, which leads to the bubble/solution layer several molecules thick consists mainly of water. The adsorption of surfactant molecules on the surface of the bubble changes the situation raising the concentration of $Na^+$ directly near the boundary. As shown in [15] the presence of any SAS in aqueous NaCl solution leads to the increase of Na emission and, interestingly, the increase is significant for nonionic SAS.

Therefore, we have got the experimental evidence for both of the models and, at the same time, there are no data that deny the possibility of emissions from both zones – the bubble/liquid boundary and the hot gas phase inside the bubble.

The correlation functions of "Na-continuum" and "continuum-Na" types presented in Fig. 5 are calculated for photon pulses coming every time from an individual bubble. It means the results concern the bubbles giving both types of the flashes. We can conclude that the emission of both Na and continuum occurs simultaneously for such a bubble and Na emits after continuum with a delay of about 0.5 ns in this case. The results are the same for both NaCl and SDS solutions and show that Na emission as well as a continuum emission relates to the peak phase of the bubble collapse when the highest temperatures and densities are achieved inside the bubble.

The autocorrelation functions of "Na-Na" and "continuum-continuum" types presented in Fig. 4 are calculated for photon pulses which can be emitted either from a single bubble or from different bubbles. Regardless of this, we can conclude that the flash of Na emission lasts longer than the flash of continuum. The Na flash is the longest for SDS solution whereas the duration of continuum flash is slightly different for investigated solutions.

In accordance with [25], peak temperature in a bubble core is about 5000 K whereas the temperature of thin liquid layer immediately surrounding the collapsing bubble reaches a value of 1900 K, meaning that thermal conditions in the layer is, in principle, conditions that are supercritical for water (647 K and 218 atm). This stage of bubble collapse extends over several nanoseconds (about 20 ns as calculated for SBSL in [26]). Then opalescence should occur: spontaneous violation and collapse of surface resulting in the bubble/solution boundary moving into the liquid phase. $Na^+$ ions, separated in the solution in the hydrated state, have not enough time to diffuse; they vibrate near the positions in which they were captured by the thermal wave. Under the supercritical conditions liquid $Na^+$ ions lose the hydration shells and reduce by capturing electrons from water molecules in the ions' shell [24]. The mechanism of excitation possibly occurs by reactions with H and OH radicals [19].

The above discussion does not deny the possibility of metal emission from gas phase inside the bubble. We want to emphasize another factor. As is known from emission spectroscopy, the high temperature is not preferable for Na emission. The intensity of a spectral line obeys well-known physics:

$$I \sim N \cdot (1-X) \cdot \exp(-E/kT) \qquad (2).$$

Here $N$ is the number of atoms, $X$ is the degree of ionization, $E$ is the excitation energy, $T$ is the temperature. The existence of optimal temperature follows from (2) under which the line intensity should be the highest. Such a temperature for Na emission is 1700 - 1900 K. The optimal temperature for Na emission is achieved in the bubble core prior to the maximum of compression or after this when the bubble needs some time to cool off. It can take a few nanoseconds. If we agree with the excitation mechanism through reactions with H and OH radicals [19], we should note that the lifetime of the radicals can be up to several milliseconds both in gas and liquid phases [27]. Thus, there is a possibility that the radicals are kept in the emission zone during the phase of expansion until the temperature drops to the optimal value.

The possible evolution of emission zones is schematically shown for one bubble cycle in Fig. 6. The scheme illustrates the obtained results. The metal and the continuum flashes almost simultaneously. The continuum flash occurs when the temperature of the bubble core reaches a maximum value, i.e. at the moment of maximum compression. The optimal conditions for metal emission are provided close to the maximum compression moment in different zones (core or bubble boundary) and the process takes longer. As a result, the metal emission lasts longer than the continuum emission.

In addition, it can be assumed that adsorbed SDS molecules even more prolong the Na flash due to the blocking of metal atoms trapped inside the bubble with nanodroplets. Hydrophobic tails of SDS molecules impede the penetration of Na into solution. The difference of Na flash duration for 0.5 M and 5 M NaCl solutions has not suitable explanation yet.

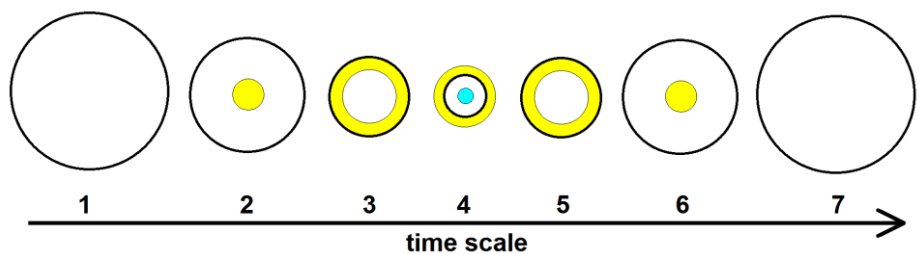

**Fig. 6.** Schematic representation of evolution of presumable emission zones for Na (yellow) and continuum (blue) close to moment of maximum bubble compression. Time points 1, 7 respond to too low temperature for emission; time points 2, 6 - conditions in bubble core are suitable for Na emission; time points 3, 5 - temperature of bubble core is high for Na emission and too low for continuum emission, however conditions on bubble/solution boundary are suitable for Na emission; time point 4 - conditions in bubble core start processes forming SL continuum, at the same time rapid heating of bubble surface moves boundary into solution, catching $Na^+$ in layer of supercritical fluid which becomes a presumable zone of emission.

## 5. Conclusions

The duration of Na and continuum flashes was determined for MBSL from Ar saturated aqueous 0.5 M NaCl, 5 M NaCl, 5mM SDS by time-correlated single photon counting using a autocorrelation between photon pulses of the same type of emission ("Na-Na" or "continuum-continuum") registered with two PMTs. The continuum flash duration turned out to be about 1.6-2.4 ns for all cases while the Na flash duration was about 4.6, 7.5 ns for NaCl solutions and about 11.2 ns for SDS solution. Using the correlation between photon pulses of Na and continuum, we detected the delay of about 0.5 ns between the flashes for all cases, with the continuum emitted first. The results suggest: a) there are the bubbles radiating both types of emission; b) the Na flash takes longer than the continuum flash, because the conditions providing the processes contributing to MBSL continuum exist for a shorter time in comparison with the conditions for Na emission; c) the duration of Na flash is longer in SDS solution, perhaps, due to the blocking of Na inside a bubble by SDS molecules adsorbed on the bubble surface.